\documentclass{article}
\usepackage{spconf,amsmath,graphicx,booktabs,multirow,url}

\title{LE-SSL-MOS: Self-Supervised Learning MOS Prediction with Listener Enhancement}
\name{Zili Qi$^1$, Xinhui Hu$^1$, Wangjin Zhou$^2$, Sheng Li$^3$, Hao Wu$^1$, Jian Lu$^1$, Xinkang Xu$^1$}
\address{
$^1$Hithink RoyalFlush AI Research Institute, Zhejiang, China \\
$^2$Kyoto University, Kyoto, Japan \\
$^3$National Institute of Information and Communications Technology (NICT), Kyoto, Japan
}
\begin{document}
%
\maketitle
\begin{abstract}
Recently, researchers have shown an increasing interest in automatically predicting the subjective evaluation for speech synthesis systems. This prediction is a challenging task, especially on the out-of-domain test set. In this paper, we proposed a novel fusion model for MOS prediction that combines supervised and unsupervised approaches. In the supervised aspect, we developed an SSL-based predictor called LE-SSL-MOS. 
The LE-SSL-MOS utilizes pre-trained self-supervised learning models and further improves prediction accuracy by utilizing the opinion scores of each utterance in the listener enhancement branch.
In the unsupervised aspect, two steps are contained: we fine-tuned the unit language model (ULM) using highly intelligible domain data to improve the correlation of an unsupervised metric - SpeechLMScore. 
Another is that we utilized ASR confidence as a new metric with the help of ensemble learning. 
To our knowledge, this is the first architecture that fuses supervised and unsupervised methods for MOS prediction.
With these approaches, our experimental results on the VoiceMOS Challenge 2023 show that LE-SSL-MOS performs better than the baseline. 
Our fusion system achieved an absolute improvement of 13\% over LE-SSL-MOS on the noisy and enhanced speech track. Our system ranked 1st and 2nd, respectively, in the French speech synthesis track and the challenge's noisy and enhanced speech track. 
\end{abstract}
\section{Introduction}
\label{sec:intro}
In speech synthesis, evaluating the quality of synthesized speech has relied on subjective evaluations such as Mean Opinion Scores (MOS) by human listeners. During an MOS listening test, listeners are asked to rate a speech sample's naturalness and quality based on their subjective impressions. The resulting score is divided into five levels, representing poor, bad, average, good, and excellent. 
Labeling datasets for synthesized speech can be very expensive and time-consuming, as it often involves having many listeners listen to many synthesized speech samples. Therefore, researchers have been developing methods that can automatically evaluate the quality of speech synthesis systems.

With the continuous development of deep neural network (DNN) techniques, researchers have gradually converted the evaluation task into a way to learn the mapping relationship between speech and its quality score through these powerful techniques and achieved promising results. 
MOSNet, a model for predicting speech quality scores based on convolutional neural networks (CNNs) and bidirectional long short-term memory networks (BLSTMs), was proposed for speech conversion tasks \cite{MOSNet2019}. 
MBNet and LDNet were proposed to use each listener's opinion scores \cite{MBNET2021, LDNet2022}. 
With its high capability of representations in speech, self-supervised learning (SSL) was also applied to MOS predictions \cite{SSL2018}. 
It was found that fine-tuning the MOS prediction with the wav2vec2.0 model had better model generalization and accuracy \cite{Generalization2022}. 
Similar results were obtained using a multi-task learning structure with SSL model features \cite{Tian2022}.
In response to the VoiceMOS Challenge 2022 \cite{Huang2022}, many innovative model architectures based on pre-trained SSL models have been proposed, including the Fusion model, UTMOS, DDOS, and ZevoMOS,  which further improved the accuracy of the MOS predictions \cite{Fusion2022, UTMOS2022, DDOS2022, Stan2022}. 
As acquiring training data is a challenge, Maiti \textit{et al.} proposed an unsupervised evaluation metric called SpeechLMScore, which evaluates the quality of speech generation by calculating the likelihood of a speech-language model \cite{Speechlmscore2023}.
A recent study on a dataset with over 1 million ratings found that models trained with data from multiple locales consistently outperformed the baseline for mono-locale data \cite{Sellam2023}.

This paper proposes a novel fusion architecture that improves estimation accuracy by utilizing supervised and unsupervised subsystems through ensemble learning.
In supervised subsystems, we employ a novel SSL-based MOS predictor, LE-SSL-MOS, by leveraging the listener-enhanced branch to learn independent opinion scores of listeners and then improving the prediction performance using a shared encoder by multi-task learning.
We utilized two unsupervised metrics, finetuned-ULM SpeechLMScore, and ASR Confidence, in unsupervised subsystems. 
The former evaluates the perplexity of speech through a fine-tuned unit language model by the highly intelligible target domain data.
The latter estimates the machine's opinion score by calculating the average probability of recognized tokens.

\section{Related Works}
\label{sec:related}
SSL speech models, such as wav2vec2.0, can learn meaningful abstract representations of speech from a large corpus of unlabeled speech data.

The results of the VoiceMOS Challenge 2022 indicate that top-ranking teams in the main track have all used methods to fine-tune SSL models, and it was observed that SSL-based models showed overwhelming effectiveness for this task \cite{Huang2022}.

Among the effective SSL-based approaches, a so-called SSL-MOS \footnote{\url{https://github.com/nii-yamagishilab/mos-finetune-ssl}} \cite{Cooper2021GeneralizationAO} is actively utilized for MOS prediction and achieved good performances in the Challenge.

\begin{figure}[htb]

\begin{minipage}[b]{1.0\linewidth}
  \centering
  \centerline{\includegraphics[width=8.5cm]{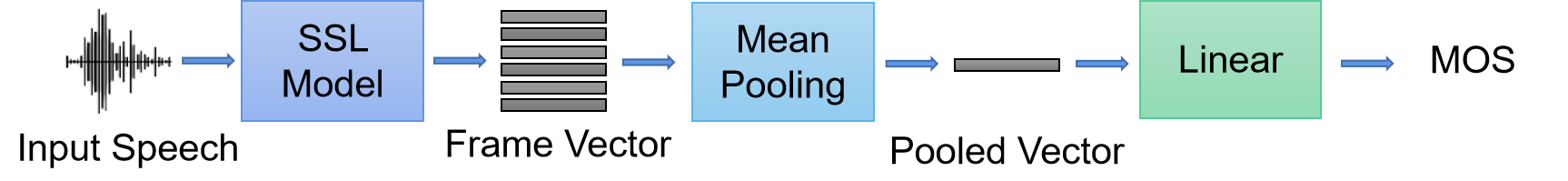}}
\end{minipage}
\caption{Block diagram of SSL-MOS approach.}
\label{fig:SSL-MOS}
\end{figure}

Therefore, we adopted the SSL-based approach for MOS prediction in this study and took the SSL-MOS as our baseline system. The block diagram in Fig.\ref{fig:SSL-MOS} shows the structure of the SSL-MOS system. Here, We will give a brief description of it.

Suppose we have a MOS dataset with \(N\) speech samples, denoted as \(\boldsymbol{X}=\{\boldsymbol{x_1, x_2, ..., x_i, ..., x_N}\}\), with \(L\) listeners, each speech sample has been rated by \(M\) listeners with \(M\) opinion scores.
To use the average opinion score during training, for each speech sample \(\boldsymbol{x_i}\), we need to first calculate the average of \(M\) independent opinion scores as the average opinion score \(y_i\) for this sample. During training, for each speech sample \(\boldsymbol{x_i}\), we need to encode the input speech into SSL representation vectors at the frame level using an encoder $\mathcal{E}$ based on the SSL model. Then, we need to convert the frame-level representation vectors to utterance-level vectors through mean pooling $\mathcal{M}$, and finally map the utterance-level vector to the predicted speech quality score \(\hat{y_i}\) through a linear layer $\mathcal{H}_{ssl}$ \cite{Generalization2022}.
\begin{align}
  \hat{y_i} &= \mathcal{H}_{ssl}(\mathcal{M}(\mathcal{E}(\boldsymbol{x_i})))
  \label{equation:eq1}
\end{align}
The goal of SSL-MOS is to estimate the utterance-level MOS, so its objective function can be defined as follows:
\begin{align}
  \mathcal{L}_{ssl} &= \frac{1}{N}\sum^{N}_{i=1}|y_i-\hat{y_i}|
  \label{equation:eq2}
\end{align}

\section{Proposed Methods}
\label{sec:proposed_methods}

In this section, we first present our proposed LE-SSL-MOS model. Next, we will formulate the fusion model for track 3 of the VocieMOS Challenge 2023. 
Finally, we present an improvement to the unsupervised metric SpeechLMScore by fine-tuning a unit language model (ULM) and comparing it with SSL-based methods in experiments.
\subsection{Listener-Enhanced SSL-MOS}
\label{ssec:le_ssl_mos}

\begin{figure}[htb]
\begin{minipage}[b]{1.0\linewidth}
  \centering
  \centerline{\includegraphics[width=8.5cm]{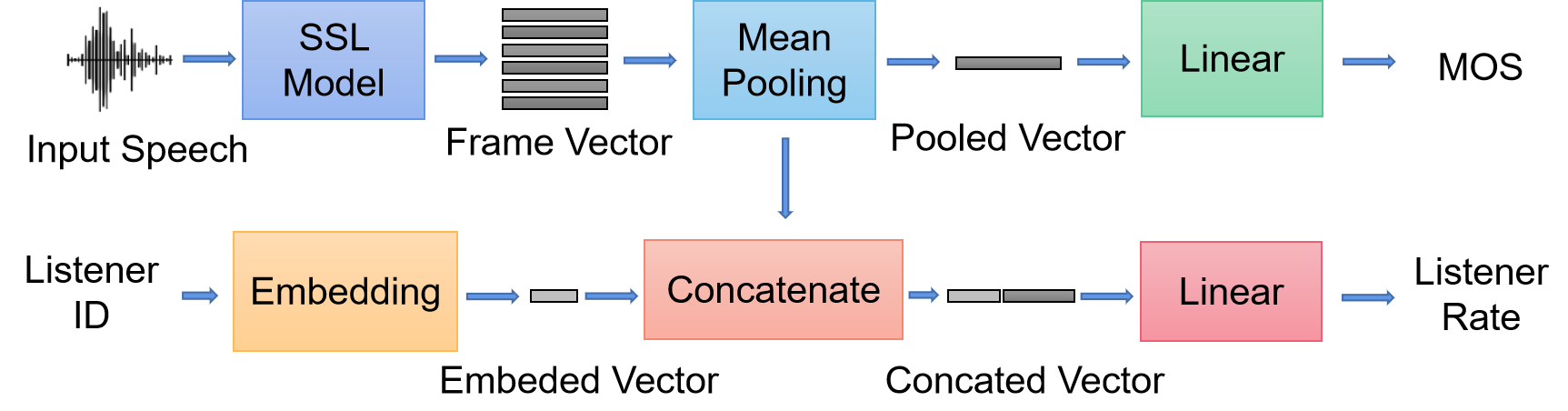}}
\end{minipage}
\caption{Architecture of the LE-SSL-MOS model}
\label{fig:LE-SSL-MOS}
\end{figure}

From Section \ref{sec:related}, we can see that SSL-MOS only utilizes the mean opinion score from the MOS dataset, while the subjective scores provided by many listeners are not fully utilized. Considering the diversity in subjective opinion distribution among each listener, we believe that improving the performance of the model by considering the opinion scores of each utterance for each listener would be beneficial. In previous studies, to make full use of the training data, LDNet (non-SSL-based method) uses the listener-dependent (LD) model to capture each listener's preferences \cite{LDNet2022}.

Following it, we propose a \textbf{L}istener-\textbf{E}nhanced SSL-MOS(LE-SSL-MOS) structure to realize that purpose, utilizing an additional branch to learn independent opinion scores of listeners, then enhance the performance of the SSL-MOS branch by using a shared encoder. its architecture shown in Fig. \ref{fig:LE-SSL-MOS}. This multi-task MOS predictor uses a shared SSL encoder to enhance the speech feature representation and two independent decoders, one for estimating MOS and the other for estimating independent opinion scores of listeners.

Compared to Fig.\ref{fig:SSL-MOS}, Fig.\ref{fig:LE-SSL-MOS} adds a listener-enhanced branch. This branch predicts the subjective opinion scores of input speech for each listener through concatenating SSL features and the embedding vector of the listener's ID. Assume there are \(M\) listeners who provide subjective scores of input speech \(\boldsymbol{x_i}\), which can be represented as \(\boldsymbol{S} = \{s_1, s_2, ..., s_M\}\). The subjective opinion score \(S_m\) of listener \(l_m\) for input speech \(\boldsymbol{x_i}\) is given by \(S_m\). During training, input speech \(\boldsymbol{x_i}\) is extracted by an encoder (SSL model) $\mathcal{E}$ and a mean pooling $\mathcal{M}$ layer to obtain utterance-level feature and the ID of listener \(l_m\) is obtained by an embedding module $\mathcal{B}$ to obtain its embedding vector. After concatenating these two vectors, they are input into a linear layer $\mathcal{H}_{le}$ to obtain the prediction of listener \(l_m\)'s opinion score \(\hat{y}_i^m\) for \(\boldsymbol{x_i}\), as shown in Equ. (\ref{equation:eq4}).

\begin{align}
  \hat{y}_i^m &= \mathcal{H}_{le}(Concat(\mathcal{M}(\mathcal{E}(\boldsymbol{x_i})), \mathcal{B}(l_m)))
  \label{equation:eq4}
\end{align}

The objective function for the Listener Enhancement(LE) branch can be defined as follows:

\begin{align}
  \mathcal{L}_{le} &= \frac{1}{MN}\sum^{N}_{i=1}\sum^{M}_{m=1}|y_i^m-\hat{y}_i^m|
  \label{equation:eq5}
\end{align}
And the loss of LE-SSL-MOS can be written as:

\begin{align}
  \mathcal{L} &= \alpha \mathcal{L}_{ssl} + \beta \mathcal{L}_{le}
  \label{equation:eq6}
\end{align}
where \(\alpha\) and \(\beta\) are hyperparameters balancing the losses.

\subsection{SpeechLMScore with fine-tuned ULM}
\label{ssec:speechlmscore}

\begin{figure}[htb!]

\begin{minipage}[b]{1.0\linewidth}
  \centering
  \centerline{\includegraphics[width=8.5cm]{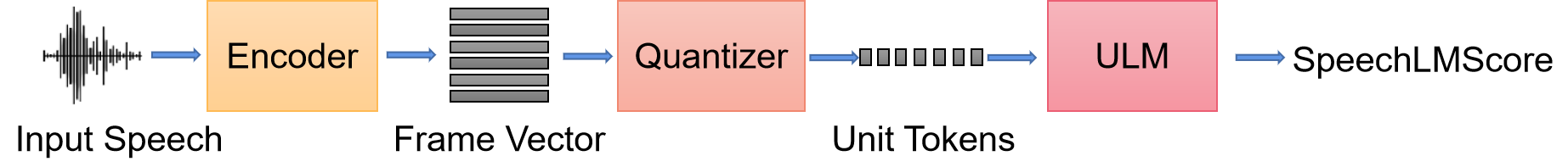}}
\end{minipage}

\caption{The block diagram of SpeechLMScore system.}
\label{fig:speechlmscore}
\end{figure}

SpeechLMscore was proposed as an extension metric for text generation to speech generation \cite{Speechlmscore2023}. The block diagram of the speechLMscore system is shown in Fig.\ref{fig:speechlmscore}.
From Fig.\ref{fig:speechlmscore}, we can see that the system consists of three parts: encoder (CPC, HuBERT or wav2vec2) \cite{CPC2021, HuBERT2021, NEURIPS2020_92d1e1eb}, quantizer (KMeans) and unit language model (ULM).
SpeechLMScore measures the average log probability of a set of unit tokens.

In natural language tasks, we often fine-tune the LM to learn the data distribution features of the target task, thereby improving its performance. 
It is natural to extend the idea of fine-tuning text-language models to speech-language models. Therefore, fine-tuning a unit language model(ULM) in the SpeechLMScore system with a small amount of data from the domain can further improve the correlation between SpeechLMScore and human subjective scoring.

\subsection{Fusion model}
\label{ssec:proposed_fusion}

Based on the work of \cite{Fusion2022}, different MOS predictors can capture different information from training sets, and ensemble learning can effectively utilize this information further to enhance the stability and accuracy of the system.
Recently, researchers have proposed many methods for self-supervised learning of speech, such as MMS, Wav2Vec2.0, XLSR, and WavLM \cite{NEURIPS2020_92d1e1eb,pratap2023scaling,XuBA22,WavLM2022}. 
In this study, we selected multiple pretrained models, for example, \(P\) models, from these methods to serve as SSL models in our SSL-based methods (SSL-MOS and LE-SSL-MOS). As a result, we can obtain \(2P\) MOS predictors based on SSL-based methods.
\cite{Chinen2022} shows that the utterance-level metrics are more useful than system-level metrics by analyzing their baseline system. 
Therefore, we selected the top \(Q\)(\(Q < 2P\)) MOS predictors under utterance-level SRCC metric as supervised subsystems for the fusion model.

\begin{figure}[ht!]
\begin{minipage}[b]{1.0\linewidth}
  \centering
  \centerline{\includegraphics[width=8.5cm]{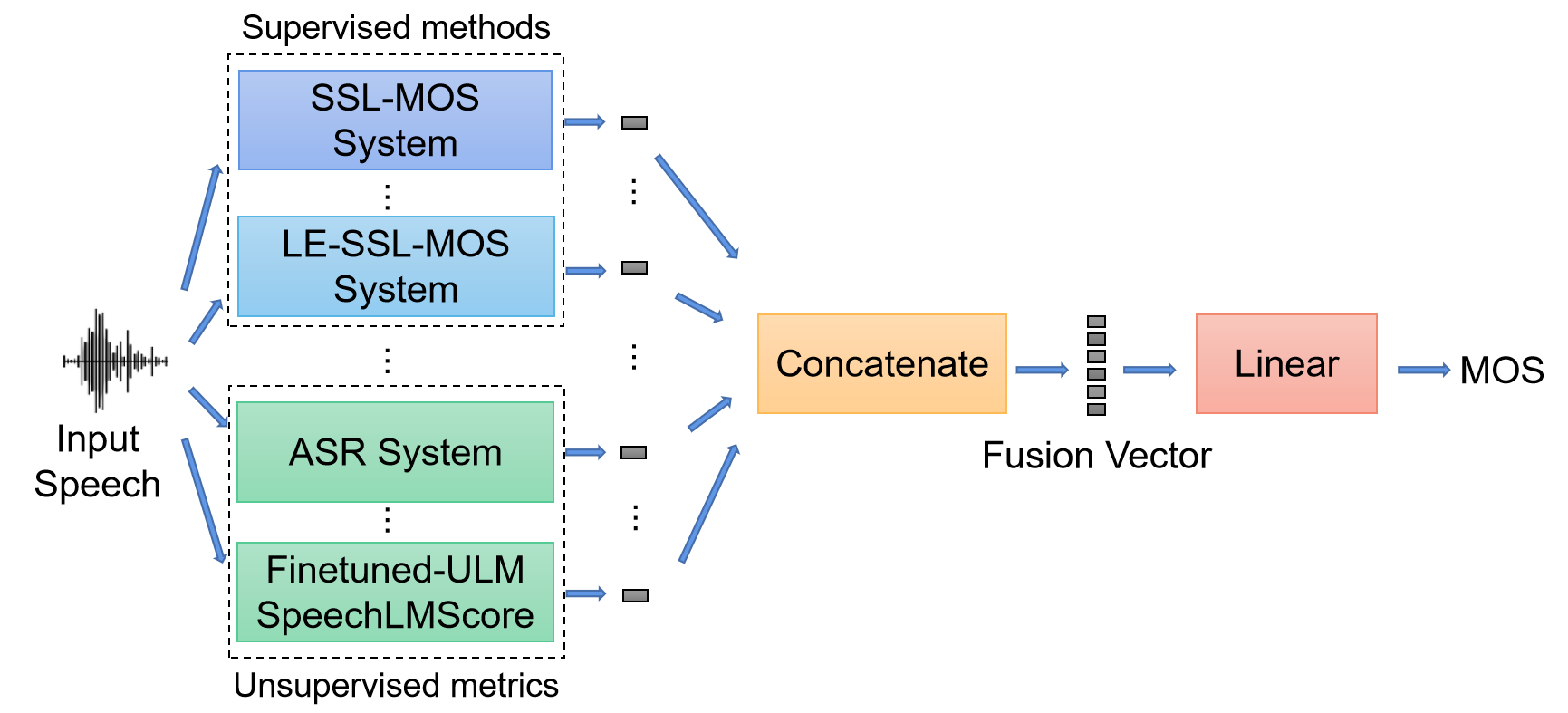}}
\end{minipage}
\caption{The system architecture of the fusion method}
\label{fig:fusion}
\end{figure}

In the Deep Noise Suppression (DNS) Challenge \cite{dubey2023icassp}, word accuracy (WAcc) is considered to be correlated with human intelligibility scores.
Additionally, we also obtained similar conclusions through the analysis of dataset TMHINT-QI \cite{chen2022inqss}, which was on a new Chinese speech dataset for speech intelligibility and quality assessment, and we found that there is a medium correlation between the average opinion score of noise-enhanced speech and recognition confidence by an ASR system. 
We believe that the speech recognition confidence by the ASR system will provide helpful information that differs from MOS prediction models for noise-enhanced speech datasets.
Therefore, we use multiple ASR systems as unsupervised subsystems of the fusion model. 
The unsupervised subsystems also contain finetuned-ULM SpeechLMScore described in section \ref{ssec:speechlmscore}.
The system framework diagram of the fusion model is shown in Fig. \ref{fig:fusion}.
Assuming that the number of ASR systems we used is \(R\) and the number of SpeechLMScore metrics is \(S\).
We will obtain our final fusion model by concatenating \((Q+R+S)\) results as input feature and using a Linear layer as a fuser. 

\section{Experiments}
\label{sec:expriments}

\begin{table*}[ht!] 
  \normalsize
  \caption{Results of SSL-based MOS predictors(SSL-MOS and LE-SSL-MOS) on the validation set of distributed TMHINT-QI}
  \label{tab:ssl_based_results}
  \centering
  \resizebox{1.0\linewidth}{!}{
  \begin{tabular}{llllllllll}
    \toprule
	\multirow{2}{*}{\textbf{SSL-based Arch}}    & \multirow{2}{*}{\textbf{SSL models}}    & \multicolumn{4}{c}{\textbf{Utterance-level}} &  \multicolumn{4}{c}{\textbf{System-level}} \\
	\cmidrule(lr){3-6}\cmidrule(lr){7-10}
	 & 	& MSE & LCC & SRCC & KTAU  & MSE & LCC & SRCC & KTAU \\
    \midrule
    \multirow{8}{*}{\textbf{SSL-MOS}}    & \textbf{MMS Base}        & 0.959    & 0.529    & 0.507    & 0.380 &  0.891    & 0.618    & 0.590    & 0.448    \\
	& MMS-1B FL102    & 1.034   & 0.475    & 0.459    & 0.341 &  0.955    & 0.550    & 0.505    & 0.391   \\
	& Wav2Vec2.0 Base   & 1.005   & 0.483    & 0.461    & 0.343 &  0.830    & 0.540    & 0.503    & 0.363   \\
	& Wav2Vec2.0 Large  & 1.004   & 0.504    & 0.484    & 0.361 &  \textbf{0.620}    & \textbf{0.682}    & \textbf{0.655}    & \textbf{0.491}   \\
	& Xlsr53 56k      & 1.087   & 0.441    & 0.426    & 0.314 &  1.120    & 0.424    & 0.374    & 0.283   \\
	& WavLM Base      & 0.991   & 0.500    & 0.473    & 0.352 &  0.872    & 0.569    & 0.536    & 0.395   \\
	& WavLM Base+     & 1.025   & 0.505    & 0.482    & 0.358 &  0.845    & 0.585    & 0.556    & 0.401   \\
	& WavLM Large     & 1.143   & 0.447    & 0.427    & 0.315 &  0.991    & 0.552    & 0.519    & 0.374   \\
	\midrule
    \multirow{8}{*}{\textbf{LE-SSL-MOS}}    & \textbf{MMS Base}        & \textbf{0.951}    & \textbf{0.537}    & \textbf{0.517}    & \textbf{0.388} &  0.875    & 0.607    & 0.566    & 0.426    \\
	& \textbf{MMS-1B FL102}    & 1.080   & 0.509    & 0.490    & 0.366 &  1.217    & 0.584    & 0.542    & 0.413   \\
	& Wav2Vec2.0 Base   & 1.026   & 0.497    & 0.473    & 0.353 &  0.967    & 0.579    & 0.551    & 0.408   \\
	& Wav2Vec2.0 Large  & 0.993   & 0.505    & 0.485    & 0.362 &  0.635    & 0.666    & 0.634    & 0.474   \\
	& XLSR53 56k      & 1.162   & 0.374    & 0.368    & 0.270 &  0.916    & 0.574    & 0.553    & 0.393   \\
	& \textbf{WavLM Base}      & 0.964   & 0.526    & 0.500    & 0.373 &  0.660    & 0.639    & 0.611    & 0.459   \\
	& \textbf{WavLM Base+}     & 0.969   & 0.529    & 0.511    & 0.382 &  0.736    & 0.620    & 0.591    & 0.435   \\
	& WavLM Large     & 0.985   & 0.496    & 0.478    & 0.355 &  0.754    & 0.638    & 0.639    & 0.474   \\
    \bottomrule
	  \end{tabular}}
\end{table*}

\subsection{Evaluation criteria}
The evaluation criteria include mean squared error (MSE), Linear Correlation Coefficient (LCC), Spearman Rank Correlation Coefficient (SRCC), and Kendall Tau Rank Correlation (KTAU) at the utterance level and system level \cite{MOSNet2019}. 
MSE measures the difference between the predicted and ground-truth scores, while the other criteria evaluate the correlation.

\subsection{Training phase}
\label{ssec:train_phase}

This section provides detailed information about the experiments and results we conducted during our training phase. During the training phase, the organizers of the VoiceMOS Challenge 2023 released a noise-enhanced speech dataset called TMHINT-QI \cite{chen2022inqss}. To ensure that the testing set contained systems unseen in the training set, the organizers re-split the training set (8201 utterances) and validation set (4736), with 36 systems from the validation set not appearing in the training set. Before training, we resampled all audio data to 16kHz and normalized the amplitude of speech by using the toolkit sv56 \cite{Huang2022}.

\subsubsection{SSL-based MOS Predictors}
\label{sssec:ssl_based_mos_predictors}

During training, we first selected 8 SSL models (MMS Base, MMS-1B FL102, Wav2Vec2.0 Base, Wav2Vec2.0 Large, XLSR53 56k, WavLM Base, WavLM Base+ and WavLM Large)\footnote{\url{https://github.com/facebookresearch/fairseq}} to be used for two SSL-based architectures (SSL-MOS and LE-SSL-MOS), and fine-tuned on the above datasets \cite{NEURIPS2020_92d1e1eb,pratap2023scaling,XuBA22,WavLM2022}. These models are trained for 1000 epochs, using L1-loss, batch size of 4, learning rate of 1e-4, an SGD optimizer, and stopping training early if the loss on the validation set does not decrease in 10 epochs. In addition, for the hyperparameters of LE-SSL-MOS, \(\alpha\) and \(\beta\) are both equal to 1, and the dimension of the listener ID embedding feature is equal to 128. During the prediction phase, we only keep the SSL-MOS branch for LE-SSL-MOS. Table \ref{tab:ssl_based_results} shows the results of all SSL-based models.

From Table \ref{tab:ssl_based_results}, we can find that under the same model architecture, MMS Base provides a higher correlation than other models under the Utterance-level SRCC metric. Meanwhile, when using the same SSL model, the LE-SSL-MOS architecture outperforms the baseline significantly.

\begin{table*}[ht!]
  \normalsize
  \caption{\normalsize{Results of confidence score, SpeechLMScore, and fusion model on the validation set of distributed TMHINT-QI. \(Confidence\) represent ASR system. \(SpeechLMScore_{ft-ULM}\) represent finetuned ULM SpeechLMScore metric, \(Fusion Model_{Q}\) represent mono SSL-based MOS predictors subsystems. \(Fusion Model_{QR}\) add ASR subsystems based on \(Fusion Model_{Q}\). \(Fusion Model_{QRS}\) add finetuned-ULM SpeechLMScore subsystems based on \(Fusion Model_{QR}\).}}
  \label{tab:fusion_results}
  \centering
  \resizebox{1.0\linewidth}{!}{
  \begin{tabular}{lllllllllll}
    \toprule
	\multirow{2}{*}{\textbf{Method type}}  & \multirow{2}{*}{\textbf{Models}}  & \multicolumn{4}{c}{\textbf{Utterance-level}} &  \multicolumn{4}{c}{\textbf{System-level}} \\
	\cmidrule(lr){3-6}\cmidrule(lr){7-10}
	 	& & MSE & LCC & SRCC & KTAU  & MSE & LCC & SRCC & KTAU \\
    \midrule
      \multirow{2}{*}{\textbf{Unsupervised metric}}  &  Confidence & -    & 0.405    & 0.418    & 0.308 &  -    & 0.576    & 0.558    & 0.418    \\
       & SpeechLMScore\(_{ft-ULM}\) & -    & 0.463    & 0.425    & 0.317  &  -    & 0.5837    & 0.563    & 0.424    \\
    \midrule
	\multirow{3}{*}{\textbf{Fusion model}}  &    Fusion Model\(_{Q}\) & 0.929   & 0.548    & 0.525    & 0.395 &  0.928    & 0.631    & 0.592   & 0.448   \\
	   & Fusion Model\(_{QR}\) & 0.925   & 0.550    & \textbf{0.529}    & 0.397 &  0.911    & 0.632    & 0.595    & 0.448   \\
	   & Fusion Model\(_{QRS}\) & 1.010   & 0.540    & 0.514    & 0.384 &  0.987    & 0.662    & \textbf{0.639}    & 0.472   \\
    \bottomrule
  \end{tabular}}
\end{table*}

\begin{table*}[ht!]
  \caption{\normalsize{System-level Results of VoiceMOS Challenge 2023 Tracks(since two isolated listener test results in track 1, we separately reported their performance, track 1a and track 1b. \(multi\) and \(track3\) respectively using the \(multi\) dataset and the distributed TMHINT-QI dataset during training. The bold and italicized parts are our final submission(T06) results in the VoiceMOS Challenge 2023.)}}
  \label{tab:voicemos_results}
  \centering
  \resizebox{0.9\linewidth}{!}{
  \begin{tabular}{lllllllll}
    \toprule
	  \multirow{2}{*}{\textbf{Models}}  & \multicolumn{4}{c}{\textbf{System-level SRCC}} & \multicolumn{4}{c}{\textbf{System-level MSE}} \\
	\cmidrule(lr){2-5}\cmidrule(lr){6-9}
	    & Track 1a & Track 1b & Track 2 & Track 3 & Track 1a & Track 1b & Track 2 & Track 3 \\
    \midrule
        SSL-MOS\(_{multi}\) & \emph{\textbf{0.790}}    & \emph{\textbf{0.750}}    & \emph{\textbf{0.693}}    & -  & \emph{\textbf{0.253}}    & \emph{\textbf{0.659}}    & \emph{\textbf{1.631}}    & -   \\
	LE-SSL-MOS\(_{multi}\) & 0.878   & 0.627    & 0.717    & - & 0.236   & 0.494    & 2.190    &  -  \\
	LE-SSL-MOS\(_{track3}\) &  -  &   -  &  -   & \emph{\textbf{0.750}}  &  -  &   -  &  -   & \emph{\textbf{0.404}}  \\
	Fusion-Model\(_{track3}\) &  -  &  -   &   -  & 0.880 &  -  &  -   &   -  & 0.206   \\
    \bottomrule
  \end{tabular}}
\end{table*}

\subsubsection{Fusion Model}
\label{sssec:fusion_model}

We use the experimental results of section \ref{sssec:ssl_based_mos_predictors} to select the MOS predictor subsystems for the fusion model. We selected the top 5(\(Q=5\)) mono-models (bold predictors in table \ref{tab:ssl_based_results}) under utterance-level SRCC metric. 
For the ASR subsystem(\(R=1\)), we use the fine-tuning model MMS-1B:FL102 based on MMS project \cite{pratap2023scaling} and compute the average log-probability of output tokens predicted by the acoustic model as the confidence. 
For the SpeechLMScore subsystem(\(S=1\)), we select librispeech clean-100 and TMHINT-QI data with MOS greater than four as the domain data. To compute SpeechLMScore, we used pre-trained HuBERT BASE as an encoder to output the third layer features as speech features and kMeans (200 clusters) as a quantizer, and a fine-tuned ULM by the above domain data \cite{fairseq2019,HuBERT2021}.

During training, we obtain the prediction results of the training set and validation set in each subsystem and concatenate them as input features for the model fuser, a fully connected layer without bias. We trained the fuser model for 1000 epochs, using MSE-Loss, batch size of 4, the learning rate of 1e-5, an RMSProp optimizer, and stopping training early if the loss on the validation set does not decrease in 20 epochs.

Compared to the best mono-model in Table \ref{tab:ssl_based_results}, the results of \(Fusion Model_{Q}\) show that utilizing ensemble learning from SSL-base MOS predictors can achieve higher utterance-level correlation. 
Table \ref{tab:fusion_results} also indicates that the fusion model with recognition confidence(\(Fusion Model_{QR}\)) can further improve performance than only used MOS predictors. And the fusion model with all subsystems(\(Fusion Model_{QRS}\)) provided the highest system-level SRCC.

\subsection{Evaluation phase}
\label{ssec:eval_phase}

This section will describe the experiment setups and results for the three released tracks in VoiceMOS Challenge2023 using SSL-based approaches and a fusion model. 

As shown in Table \ref{tab:ssl_based_results}, we selected MMS BASE as the SSL model in the SSL-based evaluation predictor. 

For track1 and track2, since the organizers did not provide training and validation sets, we collected SOMOS and the main track data from the VoiceMOS challenge 2022 and combined the training and validation sets from both datasets as the training and validation set for these tasks \cite{Huang2022,maniati22_interspeech}. We named this dataset as \emph{multi}. Using the same training method and parameters in section \ref{sssec:ssl_based_mos_predictors} and train data \emph{multi}, we obtained the final evaluation models SSL-MOS\(_{multi}\) and LE-SSL-MOS\(_{multi}\).
For track3, we directly use the MOS predictor trained in section \ref{ssec:train_phase}, and select LE-SSL-MOS(with MMS BASE in section \ref{sssec:ssl_based_mos_predictors}) and the fusion model(Fusion Model\(_{QRS}\) in section \ref{sssec:fusion_model}) as our final MOS predictor.

The system-level SRCC and system-level MSE of the challenge dataset are shown in Table \ref{tab:voicemos_results}. 

Our submission ranked 1st and 2nd, respectively, in track1a and track3 of the challenge. 
The final fusion model achieved an absolute improvement of 13\% over mono LE-SSL-MOS on track 3.

\section{Conclusions}
\label{sec:conclusions}

In this work, we propose a novel MOS predictor, LE-SSL-MOS, which constructs the predictor by utilizing the pre-trained speech self-supervised learning model, the MOS of each utterance, and the independent opinion scores of the listener. 
We will also use ensemble learning to fuse mono SSL-based MOS predictors, ASR systems, and fine-tuned-ULM SpeechLMScore metric to further improve the performance of the fusion system on noisy enhanced datasets.
In the future, we will further explore the possibility of achieving a general-purpose MOS predictor by constructing a more extensive dataset and a robust architecture.


\bibliographystyle{IEEEbib}
\bibliography{refs}

\end{document}